\documentstyle[12pt,epsf,psfig]{article}

\newcommand{\be}{\begin{equation}}
\newcommand{\ee}{\end{equation}}
\def\reff#1{(\ref{#1})}

\begin{document}

\title{{\vspace{-0cm} \normalsize
\hfill \parbox{40mm}{CERN-TH/98-110}}\\[40mm]
Comparison of Improved and Unimproved \\ Quenched Hadron Spectroscopy}

\author{A. Cucchieri$^a$,      
           T. Mendes$^a$, R. Petronzio$^{a,b}$ \\[1mm]
        {\small$^a$\em Dipartimento di Fisica, 
                 Universit\`a di Roma ``Tor Vergata''} \\[-1mm]
           {\small \em  and INFN, Sezione di Roma II } \\[-1mm]
        {\small \em Via della Ricerca Scientifica 1, 00133 Rome, Italy}\\
        {\small$^b$\em CERN, Theory Division, CH-1211 Geneva 23,
          Switzerland}
        }

\date{}
\maketitle

\begin{abstract}
We make a comparison between our quenched-hadron-spectroscopy results 
for the non-perturbatively-improved Wilson action and
the corresponding unimproved case, at $\beta=6.2$ on the same set of gauge
configurations.
Within our statistics, we find a sizeable improvement for the baryon
spectrum and for the determination of the strange-quark mass .
\end{abstract}

\vfill
\begin{flushleft}
\begin{minipage}[t]{5. cm}
  { CERN-TH/98-110}\\
April 1998
\end{minipage}
\end{flushleft}

\thispagestyle{empty}
\clearpage

The computation cost of the extrapolation to the continuum limit of
lattice QCD simulations can be significantly reduced by using
improved actions, where the leading cutoff effects are cancelled by
suitable counterterms. It has been shown that  on-shell improvement of O(a)
is achieved by adding to the usual Wilson action the clover term, 
with a coefficient that has been determined non-perturbatively
by the ALPHA collaboration \cite{Luscher}.
 
We have studied the influence of considering this non-perturbatively-improved
 Wilson action with respect to the usual, unimproved, action.
A detailed analysis of our results for the improved case has appeared 
elsewhere \cite{spectrum}, and here we concentrate on the comparison 
between the two cases (a preliminary study can be found in
\cite{lattice}). We refer to \cite{Luscher} for the description
of the improvement programme.

We consider a lattice of volume $24^3\times 48$ and coupling $\beta = 6.2$.
We choose the following values for the hopping parameter $\kappa$ in the
unimproved case:
0.1350, 0.1400, 0.1450, 0.1506, 0.1510, 0.1517, 0.1526
(for the improved case we used:
0.1240, 0.1275, 0.1310, 0.1340, 0.1345, 0.1350, 0.1352).
We consider, for the improved as well as for the unimproved case,
all nondegenerate flavor combinations from the different
values of $\kappa$.
Our simulations were carried out on the 512-processor
computer of the APE100 series at the University of Rome ``Tor Vergata''.

Our statistics come from 104 quenched gauge configurations,
generated by a hybrid over-relaxation algorithm,
with each update corresponding to a heat-bath sweep followed by three 
over-relaxation sweeps. 
The configurations are separated by 1000 updates.
The numerical inversion of the propagator for the improved case
is described in \cite{spectrum}. For the unimproved case we have
performed the inversion in a similar manner, but for the last part
of our configurations we have implemented the SSOR algorithm 
\cite{SSOR}, which corresponds roughly to a gain of a factor 4 in the
inversion time.

The analysis of the data for the improved case is described in
\cite{spectrum}. For the unimproved case, we have followed the same
procedure, namely hadron masses are obtained from single-mass fits to the
large time behaviour of zero-momentum hadron correlators, and
the errors are estimated through a single-elimination jack-knife procedure.

We remark that we determine $\kappa_c$ for both cases by using the 
so-called Ward identity mass $m_{WI}$, considered in \cite{spectrum}
(in the unimproved case we set the coefficient $c_A$ to zero). We thus
obtain for the unimproved case the value 
\begin{equation}
\kappa_c = 0.153230(9) \quad (\mbox{from} \; m_{WI} = 0).\nonumber
\end{equation}
This may be compared with the value $\kappa_c = 0.153291(15)$, obtained from 
$M_{PS}^2 = 0$.

For the fits to the dependence of hadron masses
upon quark masses we use in the unimproved case
the bare quark mass, defined as
$ m_q(\kappa) \,\equiv\, (1/\kappa\,-\,1/{\kappa_c})/2$,
while for the improved case we use  an improved
bare quark mass \cite{Luscher} defined by
\be
{\widetilde m_q}(\kappa) \;\equiv\; m_q(\kappa)\,[1\,+\,b_m\,m_q(\kappa)]
\;\mbox{.}
\label{eq:mR}
\ee
(Note that ${\widetilde m_q}$ is the renormalized mass with $\,Z_m=1\,$.)
The improvement coefficient $b_m$ has been determined non-perturbatively
\cite{Giulia} to be  $ b_m = -0.62(3)$.
For non-degenerate-flavour cases, we use symmetric averages
of the masses defined above.

\vskip 3mm
We use physical inputs at the strange quark mass for the lattice spacing.
This avoids the inclusion of systematic uncertainties deriving from a
chiral extrapolation. The value of the strange quark mass in lattice
units is obtained by using a fit to the ratio 
$M_{PS}^2/M_V^2$ at the experimental value $(M_K/M_{K^{*}})^2$.
We get for the unimproved case the value:
\begin{equation}
m_s \;=\; 0.0316(47) \;,
\label{eq:ms}
\end{equation}
to be compared with the improved case \cite{spectrum}:
$\,m_s\,=\,0.0315(45)\,$.

The inverse lattice spacing $a^{-1}$ in MeV is then obtained by normalizing
the $K^{*}$ mass to its experimental value. We get: 
\begin{equation}
 a^{-1}_{K^{*}} \;=\; \frac{M_{K^{*}}}{M_V(m_s/2)}
 \;=\; 2943(100) \;\;\mbox{MeV,}
\end{equation}
to be compared with the result obtained in the improved case with the same
definition \cite{spectrum}:
$a^{-1}_{K^{*}} \,=\, 2561(100)\,$.
[The values for $a^{-1}$ coming from $M_{\rho}$ (the chiral value)
and $M_{\phi}$ are obtained similarly, by linear extrapolation/interpolation.]

We show in Table \ref{tab:am1_comparison} the values of the various
ratios of different determinations of $a^{-1}$ and their jack-knife
errors. The difference is between the unimproved value
and the improved case. The jack-knife errors of the differences were
computed as follows: we consider these differences for each
cluster of configurations (obtained by single elimination), and we then 
evaluate their jack-knife errors.
A distinctive difference between the two cases cannot 
be observed within our statistics, but for the ratio $r_1$, whose deviation
from unity in the improved case is more significant than in the
unimproved one.

\begin{table}[htb]
\addtolength{\tabcolsep}{1mm}
\begin{center}
\begin{tabular}{cccc}
\hline
   &  unimproved    & improved & difference \\
\hline
$r_1\equiv a^{-1}_{K^{*}}/a^{-1}_{\rho}$ & 
1.037(23) & 1.030(33) & 0.007(27) \\
\hline
$r_2\equiv  a^{-1}_{\phi}/a^{-1}_{K^{*}}$ & 
1.030(31) & 1.029(27) & 0.0008(270) \\
\hline
$r_3\equiv  a^{-1}_{\phi}/a^{-1}_{\rho} $ & 
1.067(56) & 1.060(62) & 0.008(54) \\
\hline
\end{tabular}
\caption{\label{tab:am1_comparison}
Ratios of different determinations of the inverse lattice spacing 
$a^{-1}$ 
}
\end{center}
\end{table}
\normalsize

\begin{figure}
\centerline{\psfig{figure=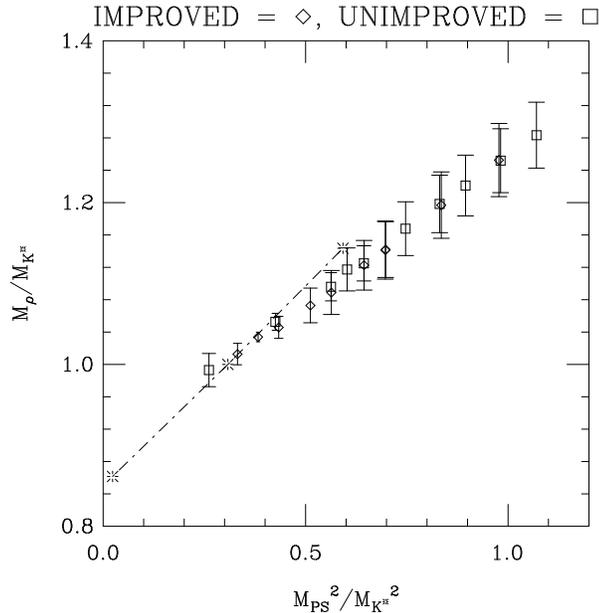,height=3.5in}}
\caption{APE plot for $M_V$. The improved case ($\Diamond$) is compared
with the unimproved one ($\Box$).}
\label{fig:APE_rho}
\end{figure}

In Fig.\ \ref{fig:APE_rho} we show a plot of the vector-meson
mass as a function of $M_{PS}^2$ for the two cases.
The masses are normalized to the $K^{*}$-meson mass, taken at the strange-quark 
mass value given in eq.\ \reff{eq:ms} above. The asterisks
correspond to the experimental values, i.e. the vector-meson masses 
$\,M_{\rho}\,$, $\,M_{K^{*}}\,$, $\,M_{\phi}\,$, and appropriate combinations
of pseudoscalar-meson masses.
The errors in the plot are calculated from independent jack-knife
procedures in the two cases.
The difference between improved and unimproved simulations can be made
statistically significant only if a jack-knife procedure is
applied to the parameters of a fit.
Indeed, the values of kappa where the inversions were taken
did correspond to different quark masses and a direct jack-knife
procedure for the hadron masses is not possible.

As can be seen from Fig.\ \ref{fig:APE_rho}, the experimental slope 
$\,dM_V/dM_{PS}^2\,$ is different from the one corresponding to our data
(this happens for the improved as well as for the unimproved case).
This slope is related to the behaviour of the dimensionless
quantity $J$ defined by:
\begin{equation}
J \;\equiv\; M_V\,(dM_V/dM_{PS}^2)
\;,
\end{equation}
obtained from $M_V$ as a function of $M_{PS}^2$.
A comparison of the $J$ values for the two cases is made in Table \ref{tab:J}.
Again, the difference was computed for each cluster, and only then was
its jack-knife error evaluated.
We see that the improvement on $J$ is marginal, which seems to indicate 
that the discrepancy between the lattice data and the experimental values
should come from the quenching approximation.
 
\begin{table}[htb]
\addtolength{\tabcolsep}{1mm}
\begin{center}
\begin{tabular}{ccccc}
\hline
 & exp. & unimproved  & improved & difference \\
\hline
$ J_{K^{*}}$ &  0.487 & 0.382(48) & 0.391(66) & -0.0081(525) \\
\hline
$ J_{\phi} $ &  0.557 & 0.409(44) & 0.427(54) & -0.019(44)   \\
\hline
\end{tabular}
\caption{\label{tab:J}
Comparison of values of the quantity $J$
}
\end{center}
\end{table}
\normalsize

\vskip 3mm
We also consider the charm-quark mass $m_c$, which we obtain from a fit
to the ratio $M_{PS}^2/M_{K^{*}}^2$ 
(the experimental input $M_D^2/M_{K^{*}}^2$ is used). 
We show in Table \ref{tab:mc_ms_comparison} a comparison of the 
ratio $m_c/m_s$ between the improved and the unimproved case. Note
that this is also the ratio of renormalized quark masses,
since in the ratio of charm over strange quark mass the renormalization
factor $Z_m$ drops out.

\begin{table}[htb]
\addtolength{\tabcolsep}{1mm}
\begin{center}
\begin{tabular}{cccc}
\hline
   &  unimproved    & improved & difference \\
\hline
$m_c/m_s$ & 12.58(190) & 10.11(140) & 2.47(140) \\
\hline
\end{tabular}
\caption{\label{tab:mc_ms_comparison}
Our lattice data for the ratio of charm-quark over strange-quark masses}
\end{center}
\end{table}
\normalsize
 
\vskip 3mm
The value of the lattice strange-quark mass given in eq.\ \reff{eq:ms} 
and the ratio $m_c/m_s$ in Table \ref{tab:mc_ms_comparison}  can be used 
to obtain two independent determinations of the physical mass of the
strange quark, as explained in \cite{spectrum}. 
The two resulting values are expected to agree,
provided that they are evolved through the
renormalization-group equation to the same scale. We use 
the conventional scale of 2 GeV.
This gives, in the unimproved case:
\begin{eqnarray}
m_s^{(1)} \;=\; 115(13) \;\mbox{MeV}  & & \mbox{from} \;m_s \\
m_s^{(2)} \;=\; 87(13) \;\mbox{MeV}  & & \mbox{from} \; m_c/m_s
\end{eqnarray}
while for the improved case 
(using the improved bare quark mass ${\widetilde m_q}$)
we obtained \cite{spectrum}:
$\,m_s^{(1)} \,=\,111(15)\,$ MeV and 
$\,m_s^{(2)} \,=\,111(16)\,$ MeV.

In order to compare the improved and unimproved cases, we consider the 
dimensionless ratio $m_s^{(1)}/m_s^{(2)}\,$.
Our results are shown in Table \ref{tab:m1_m2_comparison}. We see a
clear improvement here. This comes from two different sources:
the variation of the lattice spacing (which influences the determination 
of $\,m_s^{(1)}\,$), and the use of the improved bare quark mass
(which mainly affects the determination of $m_s^{(2)}$).

\begin{table}[htb]
\addtolength{\tabcolsep}{1mm}
\begin{center}
\begin{tabular}{cccc}
\hline
   &  unimproved    & improved & difference \\
\hline
$m_s^{(1)}/m_s^{(2)}$ & 1.315(56) & 1.005(49) & 0.310(48)  \\
\hline
\end{tabular}
\caption{\label{tab:m1_m2_comparison}
The ratio of the two strange-quark mass determinations 
}
\end{center}
\end{table}
\normalsize
 
\vskip 3mm
We now turn to the comparison of  baryonic observables.
We consider non-degenerate flavour combinations, and therefore we can 
draw APE plots with several  values of the hadron mass.
Plots are shown in Fig.\ \ref{fig:APE} for the nucleon
(or more precisely the $\Sigma$-like octet baryons) and for the $\Delta$
(decuplet baryons). We refer to \cite{spectrum} for details on how our
baryon masses are obtained. 
Again, masses are normalized to the $K^{*}$ meson mass.
Experimental points in these figures (asterisks)
correspond to $M_N$, $M_{\Sigma}$ and $M_{\Xi}$
and appropriate meson masses in Fig.\ \ref{fig:APE}a, and similarly
$M_{\Delta}$, $M_{\Sigma^{*}}$,$M_{\Xi^{*}}$ and $M_{\Omega}$ for
Fig.\ \ref{fig:APE}b. The quark-mass dependence of the decuplet
is positively affected by the improvement.

\begin{figure}
\centerline{\psfig{figure=pnr_APE_comparison.ps,height=3.5in}}
\centerline{\psfig{figure=dlt_APE_comparison.ps,height=3.5in}}
\caption{~APE plot for (a) the nucleon mass (octet baryons) and
(b) the $\Delta$ mass (decuplet baryons).
The improved case ($\Diamond$) is compared with the unimproved one ($\Box$).}
\label{fig:APE}
\end{figure}

In Table \ref{tab:masses1} we give some of our baryon mass values in MeV
(for convenience, we also include the improved values already 
reported in \cite{spectrum}).
The difference between unimproved and
improved results can be taken for each cluster as was done before; we
consider dimensionless ratios formed by the baryon mass divided by
$M_{K^{*}}$. The results are shown in Table \ref{tab:bar_comparison}.
The improved results are in much better agreement
with the experimental points. 
%
\begin{table}[htb]
\addtolength{\tabcolsep}{0mm}
\begin{center}
\vspace{-0.5cm}
\begin{tabular}{cccc}
\hline
   & exp. & improved & unimproved \\
\hline
$ M_{N} $                  & $ 939      $ & $ 952(110) $ & 1054(90) \\
\hline
$ M_{\Sigma-\Lambda} $     & $ 73.7     $ & $ 70(30) $ & 45(20) \\
\hline
$ M_{\Delta} $             & $ 1232     $ & $ 1265(110) $ & 1495(125) \\
\hline
$ M_{\Delta-N} $           & $ 293      $ & $ 297(80) $ & 336(55) \\
\hline
\end{tabular}
\caption{\label{tab:masses1}
Baryon masses in MeV (inproved and unimproved) and comparison 
with experiment. The error in 
parentheses represents the statistical error, while the error due
to the determination of the inverse lattice spacing is of about 4\%}
\end{center}
\end{table}
\normalsize
\begin{table}[htb]
\addtolength{\tabcolsep}{1mm}
\begin{center}
\begin{tabular}{ccccc}
\hline
 &  exp   &  unimproved    & improved & difference \\
\hline
$M_N/M_{K^{*}}$ & 1.05 & 1.18(11) & 1.07(13) & 0.11(13) \\
\hline
$M_{\Delta}/M_{K^{*}}$ & 1.38 & 1.68(17) & 1.35(14) & 0.34(13) \\
\hline
\end{tabular}
\caption{\label{tab:bar_comparison}
Jack-knife comparison between improved and unimproved values
for the nucleon and the $\Delta$ baryons
}
\end{center}
\end{table}
\normalsize

\vskip 3mm 
The use of a non-perturbatively-improved action  leads to 
a slightly smaller spread in the value of the lattice spacing extracted
from light mesons with and without strange quarks.
Within our statistics, it does not improve dramatically the meson
spectrum, while it affects considerably the baryons quark-mass 
dependence, especially for the decuplet, and moves it into a 
better agreement with the experimental data. 
We have also found a remarkable agreement in the improved theory with 
respect to the unimproved case
between two independent determinations of the strange-quark
mass, one normalized through the lattice spacing and the other from the
value for the charm mass  extracted in the continuum from charmonium spectrum
calculations.


\begin{thebibliography}{999}
 
\bibitem{Luscher} M. L\"uscher et al.,
                  Nucl.\ Phys.\ {\bf B478}, 365 (1996);
                  M. L\"uscher et al.,
                  Nucl.\ Phys.\ {\bf B491}, 323 (1997).

\bibitem{spectrum} A. Cucchieri et al., {\tt hep-lat}/9711040,
                   to appear in Phys.\ Lett.\ {\bf B}.

\bibitem{lattice} T. Mendes, Nucl.\ Phys.\ {\bf B} (Proc.\ Suppl.) 
                  {\bf 63}, 170 (1998).

\bibitem{SSOR} S. Fischer et al., 
               Comput.\ Phys.\ Commun.\ {\bf 98}, 20 (1996).

\bibitem{Giulia} G.M. de Divitiis and R. Petronzio,
                 {\tt hep-lat}/9710071,
                 to appear in Phys.\ Lett.\ {\bf B}.

\end{thebibliography}
\end{document}